\documentclass[twocolumn,prb]{revtex4}
\usepackage{amssymb}
\usepackage{graphicx}
\usepackage{epsfig}
\usepackage{dcolumn}
\usepackage{bm}
\usepackage{amsmath}

\begin{document}
\title{Bistable non-volatile elastic membrane memcapacitor exhibiting chaotic behavior}
\author{Julian~Martinez-Rincon}
\affiliation{Department of Physics and Astronomy and USC Nanocenter,
University of South Carolina, Columbia, South Carolina 29208,
USA}

\author{and Yuriy~V.~Pershin}
\email{pershin@physics.sc.edu}
\affiliation{Department of Physics and Astronomy and USC Nanocenter,
University of South Carolina, Columbia, South Carolina 29208,
USA}

\begin{abstract}
We suggest a realization of a bistable non-volatile memory
capacitor (memcapacitor). Its design utilizes a strained elastic
membrane as a plate of a parallel-plate capacitor. The applied
stress generates low and high capacitance configurations of the
system. We demonstrate that a voltage pulse of an appropriate
amplitude can be used to reliably switch the memcapacitor into the
desired capacitance state. Moreover, charged-voltage and
capacitance-voltage curves of such a system demonstrate hysteresis
and transition into a chaotic regime in a certain range of ac
voltage amplitudes and frequencies. Membrane memcapacitor
connected to a voltage source comprises a single element
nonautonomous chaotic circuit.
\end{abstract}

\maketitle
\section{Introduction}
Currently, there is a strong interest in
resistive, capacitive and inductive elements with memory
\cite{diventra09a}. The memory feature extends functionality of
such elements and leads to novel circuit applications
\cite{pershin09b,snider08a}. The main recent progress in this area
is in the field of memory resistive (memristive) systems
\cite{chua71a,chua76a,strukov08a} that can be potentially used in
both digital \cite{Strukov09c} and analog
\cite{pershin09b,jo10a,pershin10c,snider08a,pershin09d}
applications.

This article concerns less studied memcapacitive systems. By
definition \cite{diventra09a}, a voltage-controlled memcapacitive
system is given by the equations
\begin{eqnarray}
q(t)&=&C\left(x,V_C,t \right)V_C(t) \label{Ceq1} \\
\dot{x}&=&f\left( x,V_C,t\right) \label{Ceq2}
\end{eqnarray}
where $q(t)$ is the charge on the capacitor at time $t$, $V_C(t)$
is the applied voltage, $C$ is the {\it memcapacitance}, $x$ is a
set of $n$ state variables describing the internal state of the
system, and $f$ is a continuous $n$-dimensional vector function.
It is important that the memcapacitance $C$ depends on the state
of the system and can vary in time. Several systems showing
memcapacitive behavior has been identified including vanadium
dioxide metamaterials \cite{driscoll09a}, ionic systems
\cite{Lai09a,krems2010a} and superlattice memcapacitors
\cite{martinez09a}.

\begin{figure}[b]
\centering
\includegraphics[width=7cm]{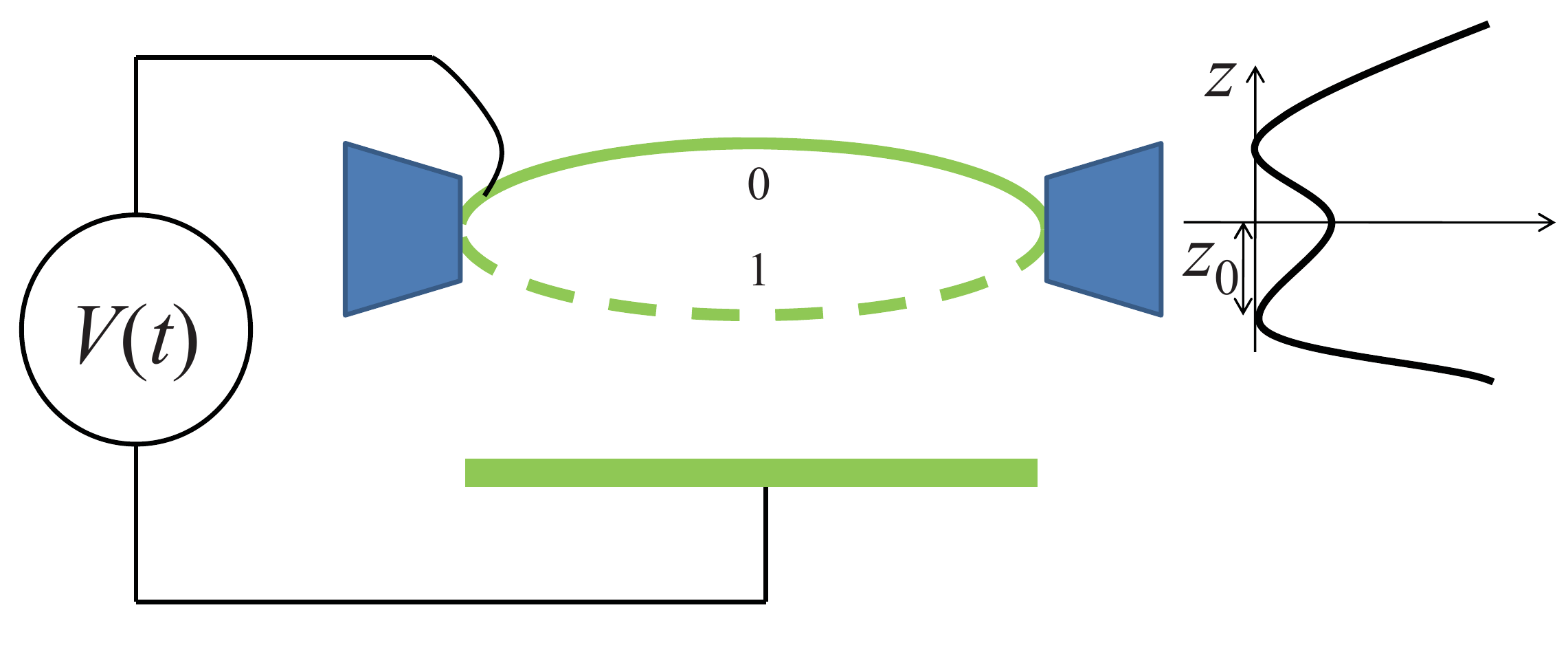}
\caption{Schematic of a bistable non-volatile elastic membrane
memcapacitor connected to a voltage source $V(t)$. The top plate
of a regular parallel-plate capacitor is replaced by a flexible
strained membrane. Because of two equilibrium positions of the
membrane (represented by a double-well potential sketched on the
right hand side of the figure), stable high and low capacitance
configurations are possible in such a system.
\label{MembraneCapacitor}}
\end{figure}

In this paper we suggest a different memcapacitor realization. Our
main idea is to replace a plate of a parallel-plate capacitor by a
strained elastic membrane as we demonstrate schematically in Fig.
\ref{MembraneCapacitor}. The applied stress bends the membrane up
or down, allowing for two equilibrium positions. When the membrane
is in the position closer to the bottom plate, the capacitance of
the device is higher (for simplicity, this configuration is called
"1"). When the membrane is bent up, the system has a lower
capacitance denoted by "0". Both states are perfectly stables,
providing a better non-volatile information storage capability
than the known designs of memcapacitors
\cite{driscoll09a,Lai09a,krems2010a,martinez09a,pershin10d}.
We show below that the information can be
written into the memcapacitor state by applying a single voltage
pulse. An interesting feature of such a device is a chaotic
behavior regime achievable within a certain range of parameters.

\begin{figure*}[tb]
\centering
\includegraphics[width=17cm]{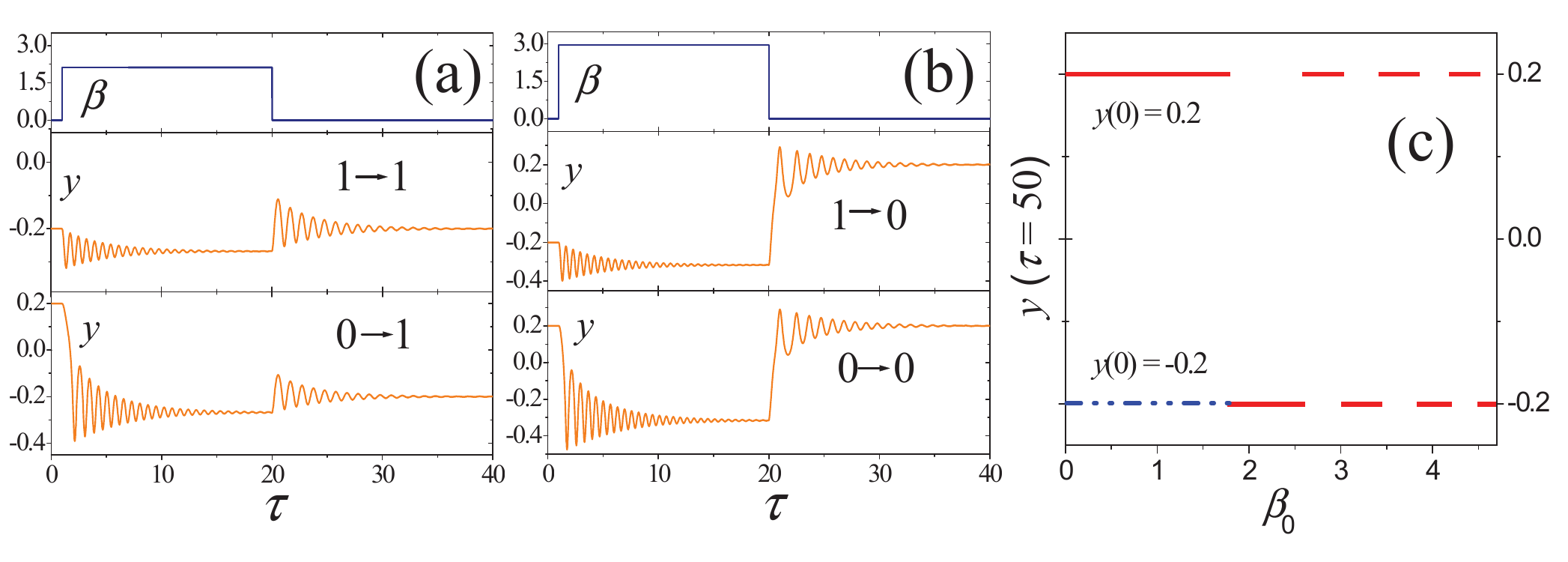}
\caption{Dynamics of membrane memcapacitor subjected to a single
voltage pulse. It is shown in (a) ((b)) that a $\beta_0=2.1$
($\beta_0=2.8$) pulse drives the system into the final high (low)
capacitance state independently of the initial state of the membrane.
The final state (at the indicated time) is shown as a function of
the pulse amplitude $\beta_0=\left[2\pi / \left( \omega_0\,d \right)\right]
\sqrt{C_0 / \left( 2\,m\right) }\,V_0$ in (c). The red (blue
double-dot-dash) curve represents the final position of the
membrane when the initial state is "0" ("1"). Below $\beta_0=1.77$
the final state depends on the initial one. Above $\beta_0=1.77$
both curves overlap and the final state is independent on the
initial state of membrane. The parameters used in these
simulations are $y_0=0.2$ and $\Gamma=0.7$.
\label{WritingMembrane}}
\end{figure*}

\section{Model} \label{model}
Let us consider an air gap capacitor composed by a fixed plate and
a flexible strained membrane (bottom and top plates in Fig.
\ref{MembraneCapacitor}). For the sake of simplicity, we describe
the membrane by a single variable $z$ (the effective displacement
of the membrane from its middle (non strained) position) and
calculate the capacitance using a parallel plates capacitor model.
The effect of the stress can be described by a double well
potential of the form $U(z)\propto\left(z^2-z_0^2\right)^2$, where
$\pm z_0$ are the equilibrium positions of the membrane. In
addition, there is an attractive electrostatic interaction between
the charges on the fixed plate and membrane. The electric force
acting on the membrane is given by the product of the charge on
it, $q$, times the electric field produced by the charge on the
fixed plate, $-q/\left(2\,\epsilon_0\,S \right)$, where
$\epsilon_0$ is the electric constant and $S$ is the area of the
fixed plate.

Introducing a dimensionless variable $y=z/d$, where $d$ is the
separation between the bottom plate and middle position of the
membrane, we write the relation between the charge $q(t)$ and
voltage $V(t)$  and the membrane's equation of motion as
%\begin{eqnarray}
\begin{equation}
q(t)=\frac{C_0}{1+y(t)}V(t)\equiv C(y(t)) V(t), \label{capacitance}
\end{equation}
\begin{equation}
\frac{\textnormal{d}}{\textnormal{d}\tau } \left[
\begin{array}{c} y \\ \\ \dot{y} \end{array} \right]= \left[
\begin{array}{c} \dot{y} \\
-4\pi^2\,y\,\left(\frac{y^2}{y^2_0}-1\right)-\Gamma\,\dot{y}-\frac{\beta^2(\tau)}{(1+y)^2}
\end{array} \right] \label{membraneequation}
%\end{eqnarray}
\end{equation}
where $C_0=\epsilon_0\,S/d$, $y_0=z_0/d$, $\Gamma=2\pi\,\gamma/\omega_0$, $\gamma$ is the damping
constant, $\omega_0$ is the natural angular frequency of the
system, $\beta(t)=\left[2\pi/ \left( \omega_0\,d \right)\right] \sqrt{C_0 / \left( 2\,m\right)
}\,V(t)$, $m$ is the mass of the membrane and the time derivatives in Eq. (\ref{membraneequation}) are taken with respect to the
dimensionless time $\tau=t\,\omega_0 / \left( 2\pi \right)$. Eqs.
(\ref{capacitance}) and (\ref{membraneequation}) have the general
form given by Eqs. (\ref{Ceq1}) and (\ref{Ceq2}) and, therefore,
the membrane memcapacitor is a
second-order voltage-controlled memcapacitive system.

\section{Initialization of memcapacitor state and hysteresis loops} \label{response}

A single voltage pulse of an appropriate duration and amplitude
can be used for the purpose of the memcapacitor
state initialization. Specifically, we demonstrate below that depending on the
pulse amplitude, the memcapacitor can be controllably set into the
state "0" or "1". The initialization of the memcapacitor state can
not rely on the electrostatic interaction alone because this
interaction in the capacitor is always attractive. Therefore, in
order to obtain the state "0", a higher amplitude pulse should be
applied such that the restoring elastic force, when the voltage
pulse ends, allows overcoming the potential barrier shown in Fig.
\ref{MembraneCapacitor}.

Figs. \ref{WritingMembrane}(a) and \ref{WritingMembrane}(b) show simulation results of the
memcapacitor dynamics obtained as a numerical solution of Eq.
(\ref{membraneequation}) for two different values of the applied
pulse amplitude. When a lower amplitude $\beta_0=2.1$ pulse is
applied to the system for a long enough time, the membrane first
finds an equilibrium position at this value of voltage. This position is independent on the initial state of the membrane in the switching regime ($\beta_0>1.77$ in our simulations, see Figs. \ref{WritingMembrane}(a) and \ref{WritingMembrane}(c) for details). Then, when the
voltage drops to zero, the membrane oscillates around $-y_0$ and
eventually the equilibrium position "1" is reached.

When a higher amplitude pulse ($\beta_0=2.8$) is applied (see Fig.
\ref{WritingMembrane}(b)), the membrane is attracted down stronger
before being released. As a result, when the pulse ends, the
membrane goes up and stays in the low capacitance configuration
"0". With a further increase of the pulse amplitude, the final
state of the membrane can be again "1" and then "0", etc. Such
periodic situation is clearly seen in Fig.
\ref{WritingMembrane}(c). Note that in the switching regime
($\beta_0>1.77$) the final state of the membrane does not depend
on its initial state.

\begin{figure}[bt]
\centering
\includegraphics[width=7cm]{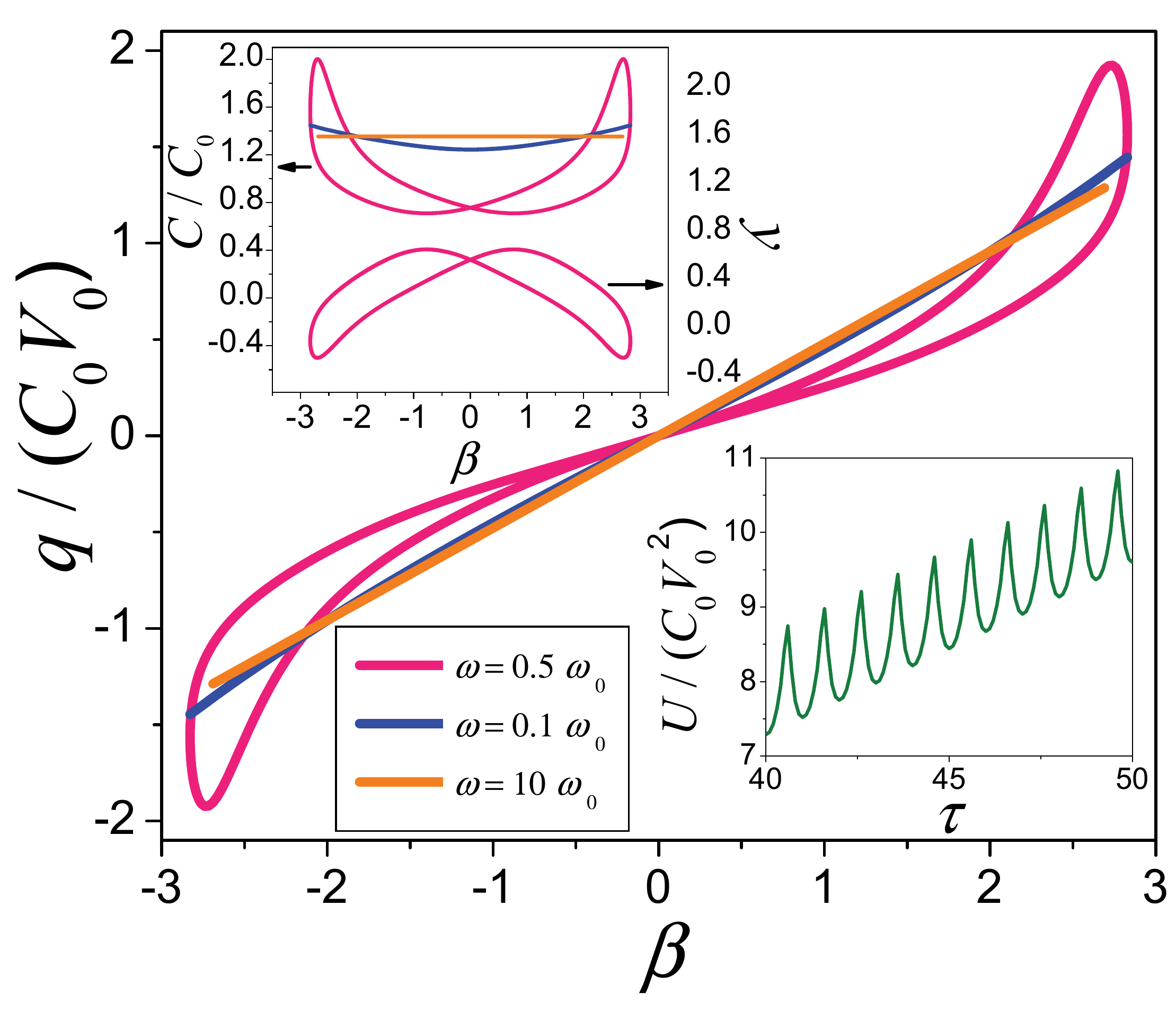}
\caption{$q-V$ and $C-V$ (top curves in the top inset) curves for a membrane memcapacitor
calculated when a sinusoidal voltage is applied. The bottom curve in the top inset represents $y(\beta)$ at $\omega=0.5 \omega_0$. Parameters used
in this simulation are the same as in Fig. \ref{WritingMembrane}
and $\beta=\beta_0\sin(\omega t)$ with $\beta_0=2.8$. The graphs
demonstrate a periodic steady-state regime beyond an initial
equilibration interval $(\tau\gtrsim30)$. Limits to linear
(orange) and non-linear (blue) behaviors are obtained at high and
low frequencies. Hysteresis loops are seen at intermediate
frequencies. The bottom inset is the energy dissipated by the
system, given by $U=\int{V\,dq}$, calculated at $\omega =0.5
\omega_0$. \label{Pinched}}
\end{figure}

When the membrane memcapacitor is subjected to ac
voltage, its dynamics can be periodic or chaotic depending on the frequency and amplitude of
the ac voltage. In this section we consider only the periodic case.
The chaotic behavior is discussed in Sec. \ref{sec_chaos}. A
numerical solution of Eqs. (\ref{capacitance},\ref{membraneequation}) with
$\beta(t)=\beta_0\sin(\omega t)$ is plotted in Fig. \ref{Pinched}
for three different values of the applied voltage frequency. The hysteresis
curves demonstrate typical behavior of memory elements
\cite{diventra09a}: non-linear dependencies at lower frequencies,
pinched hysteresis loops at intermediate frequencies and linear
behavior at higher frequencies.

The hysteresis observed in Fig. \ref{Pinched} at $\omega=0.5\omega_0$ is basically related to the well-known phase
shift between periodic applied force and position of a driven mechanical system.
Basically, this effect originates from the membrane's inertia as well as from the fact that a finite
time is required to dissipate instantaneous energy stored, e.g., in the elastic degree of freedom of the membrane.
Since the electrostatic interaction between capacitor's plates is always attractive,
the $C-V$ curve is symmetric with respect to $\beta=0$ and the membrane experiences
a driving force of the doubled frequency of ac voltage.
The pinched behavior additionally proofs that the system
is a memory capacitive device. The bottom inset in Fig. \ref{Pinched}
shows the energy dissipated by the device for 10 cycles in the
steady regime. Because of the damping term in Eq.
(\ref{membraneequation}), the membrane memcapacitor is a
dissipative system.

\section{Chaotic Behavior} \label{sec_chaos}

\begin{figure*}[tb]
\centering
\includegraphics[width=12cm]{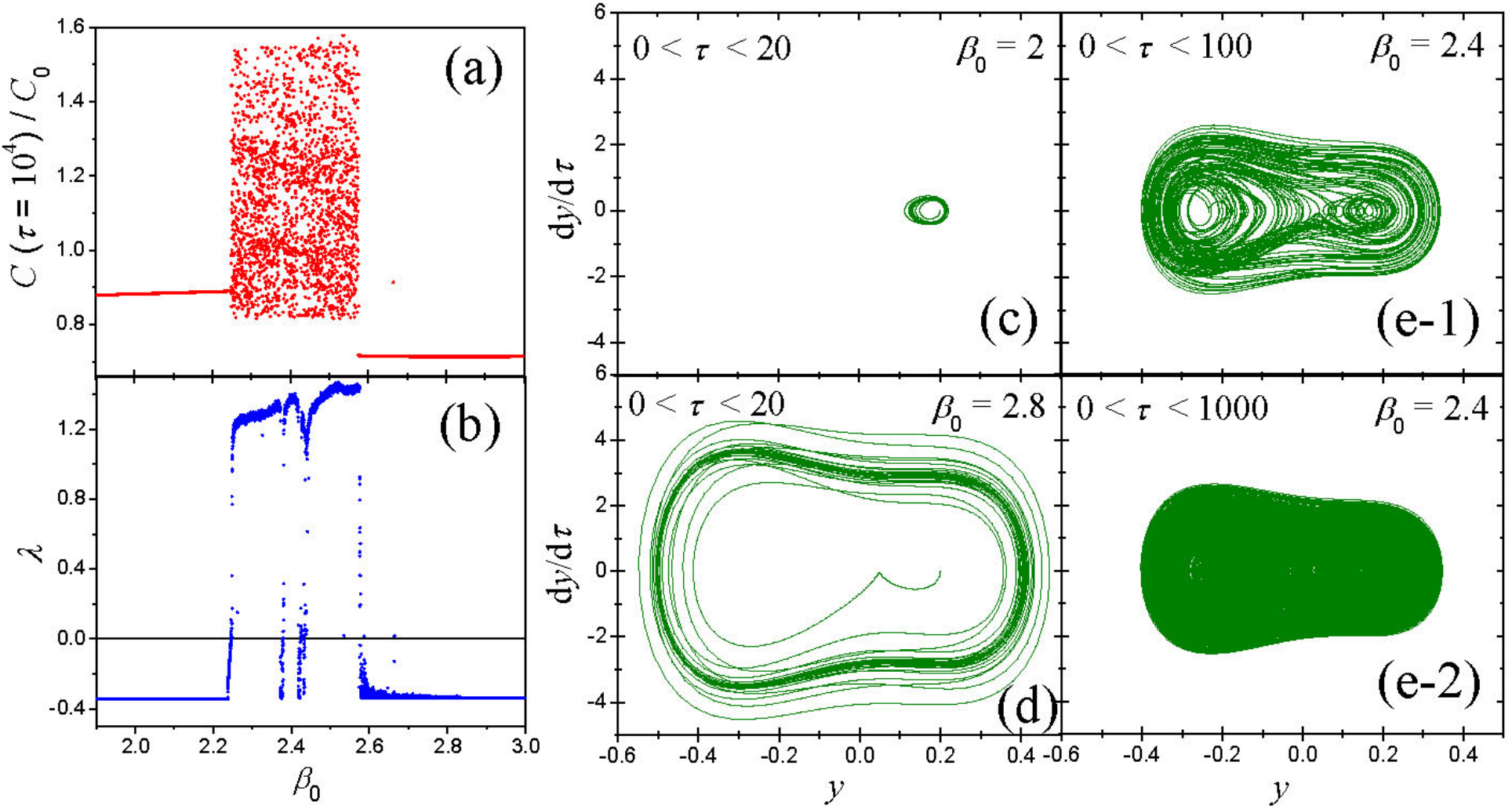}
\caption{Chaotic behavior of the membrane memcapacitor subjected
to ac-voltage ($\beta=\beta_0\sin(\omega t)$). The value of
capacitance (a) at $\tau=10^4$ and the Largest Lyapunov exponent
(b) are plotted as a function of the amplitude $\beta_0$. The
remaining plots (c)-(e) demonstrate trajectories in the phase
space ($\dot{y}$ vs. $y$) of the system at three different values
of the amplitude $\beta_0$. Plots (c) and (d) show a periodic
steady-state behavior, and (e) represents the ergodic behavior at
$\beta_0=2.4$. Calculation parameters are the same as in Fig.
\ref{WritingMembrane} and $\omega=0.5\,\omega_0$.\label{Chaos}}
\end{figure*}

Most experts would agree defining chaos as the
\textit{aperiodic, long-term behavior of a bounded, deterministic
system that exhibits sensitive dependence on initial conditions
and system parameters} \cite{Sprott03a}. In this section, we
demonstrate that the membrane memcapacitor described by equation
(\ref{membraneequation}) can exhibit chaotic behavior. In
particular, chaotic regime is observed in numerical simulations
for a specific range of amplitudes and frequencies of the applied
sinusoidal voltage $V=V_0\sin(\omega t)$.

Specifically, in order to explore chaotic properties of the
system, let us investigate how the state of the memcapacitor (at a
long selected moment of time) depends on one parameter of the
system, such as the ac-voltage amplitude $V_0$. For this purpose,
we ran multiple simulation of the system dynamics at different
values of $\beta_0$ and monitor the value of capacitance $C$
at a certain moment of time. Fig.
 \ref{Chaos}(a) shows the value of capacitance evaluated at $\tau=10^4$ as a function of the amplitude $\beta_0$. This plot demonstrates that at small
($\beta_0 < 2.25$) and high ($\beta_0 > 2.58$) voltages, the
system reaches practically the same value of capacitance (at a
given time) if the voltage amplitude is slightly changed. However,
at intermediate voltages ($2.25 < \beta_0 < 2.58$) predictability
in the value of the capacitance is lost, which is an indicator of
chaotic behavior.

The Largest Lyapunov Exponent (LLE) $\lambda$ is a quantitative
measure of chaos \cite{Sprott03a}. It is defined for a long time
behavior of a single trajectory and is sensitive to the initial
conditions of the system. If LLE is positive ($\lambda>0$) the
trajectory is chaotic and $\lambda$ is a measure of the average
rate at which predictability is lost. On the other hand, if
$\lambda\leq0$ the evolution is not chaotic. We use a standard
approach to calculate the LLE \cite{LLE,LLE1}. Fig. \ref{Chaos}(b)
shows the result of our calculation of LLE as a function of the
applied voltage amplitude $\beta_0$. The chaotic behavior
(positive values of $\lambda$) is clearly observed in the interval
$2.25<\beta_0<2.58$.

The system trajectories in the phase space ($\dot{y}$ vs. $y$) at
low, high and intermediate voltage amplitudes are shown in Fig.
\ref{Chaos}(c)-(e) correspondingly. If the voltage amplitude is
such that $\beta$ is below approximately $2.25$ (Fig.
\ref{Chaos}(c)), the membrane moves periodically either in the low
or high capacitance region. If the voltage amplitude is high
($\beta$ is above $2.58$), a different kind of periodic
steady-state solution is observed (see Fig. \ref{Chaos}(d)). In
this case, the voltage is high enough for the membrane to pass
through the barrier between the potential wells generating
oscillations of the membrane between the high and low capacitance
regions. At both low and high voltage amplitudes, the system
dynamics is not chaotic. Finally, at intermediate voltage
amplitudes ($2.25 < \beta_0 < 2.58$), the system does not reach
any steady solution. Figs. \ref{Chaos}(e-1) and \ref{Chaos}(e-2)
show the system trajectories up to two different simulation times
($\tau=100$ and $\tau=1000$). It is evident that the system is
ergodic, what is another important manifestation of chaos. In
fact, it is not quite surprising that Eq. (\ref{membraneequation})
produces chaotic trajectories. Generally, Eq.
(\ref{membraneequation}) can be seen as a modification of the
Duffing oscillator equation, whose chaotic properties are
well-known \cite{Sprott03a}.

\section{Conclusions} \label{conclusions}

The membrane memcapacitor is a modification of a parallel-plate
capacitor, in which one of the plates is replaced by a strained
membrane having two equilibrium positions. This feature defines
two states ("0" and "1") that can be used in memory applications.
The system is identified as a second-order voltage-controlled
memcapacitive system. Pinched hysteresis loops for both
capacitance and charge as a function of voltage were obtained. In
addition, a chaotic behavior was demonstrated in a certain range
of applied voltage amplitudes.

Several system parameters were used in our numerical simulations. From engineering standpoint, the
most important parameter is $z_0$ (defining the equilibrium positions of the membrane). The choice of $z_0$ should provide
well defined states "0" and "1". Moreover, a contact between the plates should be avoided at any time.
Our calculations indicate that both conditions are satisfied at $z_0=0.2d$. Smaller (larger) values of
$z_0$ (at fixed values of all other parameters) will decrease (increase) the local potential maxima at $z=0$  and thus facilitate (complicate) the mechanical switching between low- and high-capacitance states. Moreover, a variation of $z_0$ will modify the range of parameters in which the system exhibits the chaotical behavior. Basically, it is expected that at smaller (larger) values of $z_0$ chaos will be observed at smaller (large) applied voltage amplitudes. Similarly, the window of non-chaotic behavior will be modified, although the main frequency features of hysteresis loops will remain unchanged. The chaotic behavior will disappeared in the limit $z_0\rightarrow 0$ .

It is important to note that our calculations show that chaos
is possible in an ac-driven {\it single-element} electronic
circuit. Previously suggested chaotic circuits
\cite{Matsumoto84b,Chua92a,chuabook,Muthuswamy09a},
involving those with memristors \cite{Muthuswamy09a},
involve several circuit elements. Moreover, the membrane
memcapacitor is a passive device in contrast with the active
Chua's diode \cite{chuabook} and active
memristors\cite{Muthuswamy09a} used in existing chaotic
circuits. Finally, we are not aware about any other capacitive
devices exhibiting chaos. The membrane memcapacitor can possibly
be fabricated using standard MEMS or NEMS (nano-electro-mechanical
system) fabrication techniques.

\bibliography{memcapacitor1}

\end{document}